\documentclass[acmlarge,nonacm]{acmart}
\begin{document}

\title{Behind the Mask: A Taxonomic Analysis of Activities in Online Social Networks}

\author{Debora F. de Souza}
\affiliation{%
  \institution{Tallinn University}
  \city{Tallinn}
  \country{Estonia} }
\email{deboracs@tlu.ee}

\author{Gabriela Beltrão}
\affiliation{%
  \institution{Tallinn University}
  \city{Tallinn}
  \country{Estonia} } 
\email{gbeltrao@tlu.ee}

\author{Berta Chulvi}
\affiliation{%
  \institution{Universitat de València}
  \city{Valencia}
  \country{Spain}  }
\email{berta.chulvi@uv.es}

\author{Sergio D'Antonio}
\affiliation{%
  \institution{Universidad Politécnica de Madrid}
  \city{Madrid}
  \country{Spain} }
\email{sergio.dantonio@upm.es}

\author{Mehmet Gokay Ozerim}
\affiliation{%
  \institution{Yasar University}
  \city{Izmir}
  \country{Turkey} }
\email{gokay.ozerim@yasar.edu.tr}

\author{Javier Torregrosa}
\affiliation{%
  \institution{Universidad Politécnica de Madrid}
  \city{Madrid}
  \country{Spain}}
\email{franciscojavier.torregrosa@upm.es}

\author{Adrian Giron}
\affiliation{%
  \institution{Universidad Politécnica de Madrid}
  \city{Madrid}
  \country{Spain}}
\email{adrian.giron@upm.es}

\author{Angel Panizo}
\affiliation{%
  \institution{Universidad Politécnica de Madrid}
  \city{Madrid}
  \country{Spain}}
\email{angel.panizo@upm.es}

\author{Pablo Miralles González}
\affiliation{%
  \institution{Universidad Politécnica de Madrid}
  \city{Madrid}
  \country{Spain}}
\email{pablo.miralles@upm.es}

\author{Helena Liz}
\affiliation{%
  \institution{Universidad Politécnica de Madrid}
  \city{Madrid}
  \country{Spain}}
\email{helena.liz@upm.es}

\author{Javier Huertas Tato}
\affiliation{%
  \institution{Universidad Politécnica de Madrid}
  \city{Madrid}
  \country{Spain}}
\email{javier.huertas.tato@upm.es}

\author{Sonia Sousa}
\affiliation{%
  \institution{Tallinn University}
  \city{Tallinn}
  \country{Estonia}
}
\email{scs@tlu.ee}

\author{Alejandro Martín}
\affiliation{%
  \institution{Universidad Politécnica de Madrid}
  \city{Madrid}
  \country{Spain}}
\email{alejandro.martin@upm.es}

\author{Monika Maciuliene}
\affiliation{%
  \institution{Mykolas Romeris University}
  \city{Vilnius}
  \country{Lithuania}}
\email{maciuliene@mruni.eu}

\author{David Camacho}
\affiliation{%
  \institution{Universidad Politécnica de Madrid}
  \city{Madrid}
  \country{Spain}}
\email{david.camacho@upm.es}

\renewcommand{\shortauthors}{F. de Souza et al.}


\begin{abstract}
The broadcast of disinformation in online social networks (OSN) is a growing concern examined across several disciplines, including human-computer interaction (HCI). The pervasive issue has been prompting novel approaches to identify the malicious actors behind the dissemination of deceptive and fabricated content. Analyzing the characteristics and activities of these actors, we designed a taxonomy informed by collaboration with subject matter experts (SMEs) and a review of the academic literature. Our study explores how to distinguish the characteristics, activities, and strategies of malicious actors on OSN and examines how they contribute to the spread of disinformation. We describe the design process and the application of the taxonomy in a case study analyzing anti-migration discourse in social media channels, and reflect on its potential to aid researchers and practitioners in the responsible design of network systems.
\end{abstract}


\ccsdesc[500]{Information systems~Social networks}
\ccsdesc[300]{Human-centered computing~Web-based interaction}



\maketitle

\section{Introduction}

The presence of disinformation is becoming an even more pressing issue for institutional media, governments, and researchers. Amid the widespread use of Online Social Networks (OSNs), a new paradigm of social interaction and communication is emerging, as these interactive spaces simultaneously allow individuals to share thoughts, ideas, and creativity and to form social communities \cite{hasib2009threatssocialnetworks}. Nevertheless, it raises growing concerns about the spread of disinformation. 
In the same manner that OSN enables the broadcast of alerts and warnings in time-sensitive situations like natural disasters and crisis events \cite{karlova2013social}, this easy-spreading fashion also allows the circulation of disinformation orchestrated by malicious actors, as it consequently loosens up the constraints for broadcasting of unverified news \cite{shu2018userprofilesfakenews}. The alarming and pervasive proliferation of deceptive content online has ultimately affected how society receives and acts on information, which is particularly dangerous during times of active crises \cite{Rajdev2015} and election periods \cite{Anderson2020}. Unsettling global statistics also highlight growing concerns about the authenticity of online content and the erosion of trust in traditional news media \cite{Newman2023}.

Pressured by the emergence of online disinformation campaigns, researchers have attempted to develop preemptive mechanisms to identify the malicious actors behind their creation and orchestration. Nevertheless, many challenges arise in distinguishing malicious actors from other lay users in the OSN, as noted by \cite{Shahid2022}, whereas detection mechanisms are constrained \cite{Karami2021} and the complexities of agency entangled in OSN platforms are considerable \cite{patel2020human}. \cite{karlova2013social} points out that the amplification of malicious actors' influence is deeply related to digital literacy and the judgment that lay users make of what is shared and diffused. Hence, manifold challenges emerge while researchers seek to develop mechanisms to identify the activities of such individuals behind the dissemination of deceptive and fabricated content.

On the one hand, users often lack awareness of the risks associated with deceptive activity and of how to identify malicious content \cite{Tandoc2020}. On the other hand, disinformation campaigns and their sensationalized content achieve a "viral” effect by amplifying their reach through low-cost, software-controlled dissemination \cite{patel2020human, shu2020disinformation}. The viral characteristics of deceptive content and amplification factors within online networks facilitate the spread of disinformation. Analyzing the dynamics of malicious actors’ campaigns in the US elections, \cite{Matatov_2022} observed that a significant amount of deceptive and malicious content achieves high engagement within less than one day, thereby posing a substantial delay for fact-checking.

Recommendations from academia provide strategies to address the growing issue by promoting digital literacy, developing user mitigation measures, and issuing platform guidelines. Lesheer and Desai \cite{lesher2022disentangling} proposed that binary characteristics of malicious users' profiles can be used to group actor types, but cautioned that such classifications are better complemented by a hybrid approach that combines human intervention and technological tools. Likewise, Zhou et al.'s study \cite{zhou2020surveyfakenews} demonstrated that aspects of malicious actors' behavior over time (e.g., logs, messages, and follower networks) can be used to reveal the strategies they employed for social influence and engagement. However, previous studies have encountered bottlenecks in developing strategies to address such user behavior across social media and contexts. Moreover, previous research cites aspects such as the algorithms and OSN infrastructure \cite{Dourado2023}, user emotions \cite{Rastogi2022}, social and political polarization \cite{PastorGalindo2020}, network effects (influential spreaders) \cite{conti2017s}, lack of verification mechanisms \cite{Shao2018}, as well as user's perception of their own anonymity \cite{dupuis2020information} as complicating factors for the implementation of coping strategies.

This research is part of a broader project focused on developing machine-learning tools to detect and profile malicious actors on OSNs. In that pursuit, we conducted a comprehensive study that integrated subject-matter expert (SME) knowledge with evidence from prior research to identify agents involved in the spread of disinformation. The study adopted a collaborative and interdisciplinary approach, drawing on expertise in psychology, natural language processing, social network analysis, and human–computer interaction. Through a combination of literature analysis and structured expert discussions, the researchers examined the landscape and developed a classification mechanism to identify malicious actors based on their characteristics, strategies, and coordination of activities. Additionally, we conducted a case study to assess the suitability of our taxonomy in a real-world context. 

The following sections outline the study background, the stages of the collaborative workshops, the data collection approach to the literature, and the development and iterations of the taxonomy. Moreover, we conclude by presenting and reflecting on the taxonomy's application in a case study.

\section{Background and Related Work}

Facing the rising concerns about the integrity of social discourse on the internet \cite{Tandoc2020, Acker_Donovan_2019}, and the threats to society's trust in the media ecosystem \cite{Newman2023}, researchers are invested into developing mechanisms to mitigate the spread of disinformation in OSN and to cope with its adjacent issues such as distrust in institutions, journalism, and government \cite{Mourao_Robertson_2019}. Various methods have been investigated to identify OSN users based on characteristics and motivational factors of users' activity through semantic analysis \cite{Karami2021} and on users' activities across networks \cite{conti2017s}. Nevertheless, classifiers trained on keywords and expressions specific to a particular event or topic often fail to identify malicious users across contexts (\cite{shu2020disinformation}). Moreover, disinformation takes hold by deploying manipulation tactics to orchestrate inauthentic behavior, making it infeasible to detect using OSN platforms' guidelines and regulations \cite{Acker_Donovan_2019, karlova2013social}.

\subsection{Disinformation in OSN}
Disinformation is "false information, spread deliberately intending to mislead and/or deceive" \cite{zhou2020surveyfakenews}. Widely present across various OSN platforms, disinformation content may be disseminated through text, audio, images, or combinations of these modalities \cite{alam2021survey-b92, montag2021psychologytiktok}. Although clear boundaries between what constitutes misinformation and disinformation are blurred, researchers put forth definitions for disinformation that imply that the content is "fabricated or deliberately manipulated" \cite{zhou2019fakenews} to deceive the reader \cite{karlova2013social}. Such orchestrated activities spread among OSN users through the strategic delivery of deceptive content that may appear highly credible to its consumers \cite{DiazRuiz2022}. Moreover, the manipulation strategies employed by malicious actors exponentially expand the reach of deceptive content, enabling it to quickly reach a large number of users \cite{Acker_Donovan_2019}. 

Despite social media companies' efforts to curb the spread of disinformation, orchestrated campaigns exploit weaknesses and gaps in their platforms and search engines, bypassing guidelines and regulations (e.g., verification processes and gatekeeping mechanisms), creating fertile ground for malicious actors to operate. Thus, exploring methods to detect and mitigate the spread of disinformation is crucial and timely, as the landscape of communication technology infrastructures and the political economy are constantly evolving, thereby advancing the generation and dissemination of malicious content \cite{Tandoc2020, Bastick2021}.

\subsection{Malicious Actors in OSN}

In the context of OSN, malicious actors are individuals or software-based users who engage in wrongful activities by creating, publishing, and/or circulating disinformation in OSN \cite{zhou2019fakenews}. These users exhibit deceptive behaviors, ranging from posting disrespectful content to acting dishonestly or unethically, and harassing other users on the platforms \cite{Rastogi2022}. 

Factors such as emotional states (e.g., anxiety) \cite{shu2020disinformation}, social identity \cite{shu2018userprofilesfakenews}, distrust in traditional media \cite{Newman2023}, biases \& beliefs \cite{shu2020disinformation}, and selective exposure \cite{lazer2018sciencefakenews} have also been identified as potential reasons for malicious actors to get away with their reputation as a credible source of information \cite{shu2020disinformation}. For instance, malicious entities gain the trust of their victims and influence them to consume and spread disinformation through gullibility-building strategies \cite{zhou2020surveyfakenews, Bastick2021}. What primarily differentiates malicious actors from lay users is their goal of spreading fabricated content on OSNs, exploiting users' lack of tools and skills for detecting falsehood \cite{shu2020disinformation}.

\subsection{Detection mechanisms}

Previous attempts to detect malicious activities in OSNs have focused on analyzing network data and user behavior patterns. Mechanisms such as behavioral analysis (i.e., posting patterns, engagement metrics),  content analysis (i.e., sentiment analysis, fact-checking), network analysis (i.e., connection patterns, influence metrics), and cross-platform analysis (i.e., account verification), are mentioned by the works of \cite{DiazRuiz2022, latah2020detection, mbona2023classifying, Dourado2023, Shao2018}. Moreover, because the malicious spread of disinformation unfolds over time, implementing mixed detection approaches can help address multiple dimensions of malicious actors’ behavior, including activity and timing information \cite{ferrara2016rise}.
Models employing style-based approaches focus on the characteristics and motivational factors of users' activities through semantic analysis \cite{Karami2021} and on propagation-based features that examine users' activities across networks \cite{conti2017s}. However, \cite{shu2020disinformation} points out that these approaches face additional challenges, as the biases and limitations of the trained models are inherent in the keywords and expressions specific to a particular event or topic, making the classifiers event-specific. The challenge is even greater for multimodal content dissemination, as previous research indicates that such environments often lack robust verification processes and gatekeeping mechanisms, thereby facilitating the spread of disinformation across combined textual, visual, and audio modalities \cite{montag2021psychologytiktok, alam2021survey-b92}.

\subsection{Contribuition}
 In response to these challenges, our study proposes a categorization framework that focuses on the characteristics, activities, and strategies of malicious actors used to spread disinformation. Our classification provides a general, event-independent method for annotating malicious activity in OSNs. 

\section{Methodology}

\begin{figure}[h]
  \centering
  \includegraphics[width=8cm]{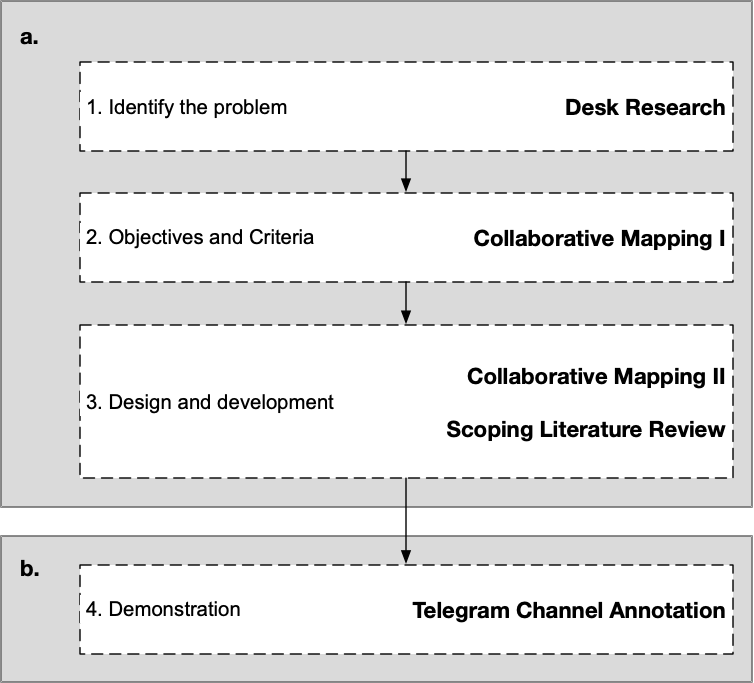}
  \caption{Details of the study design illustrating how the phases of the study are connected}
  \label{fig:study}
  \Description{Details of the study design illustrating how the phases of the study are connected}
\end{figure}

To address this issue, we designed the study following taxonomy development frameworks previously applied in ICT research \cite{Kundisch2021, nickerson2013methodtaxonomy}, as illustrated in Figure~\ref{fig:study}.
Adopting a combined inductive and deductive approach, we derived the concepts for the taxonomy through a review of pertinent literature, structured group discussions, and collaborative mapping activities conducted with subject matter experts (SMEs). Following the taxonomy development phase, we examined its application to annotating malicious content retrieved from online social networks. We focused on platforms that support the detection of multimodal content (e.g., Telegram). Our research addresses the following questions:
\begin{itemize}
\item RQ1. Who are the malicious actors?
\item RQ2. What do malicious actors do?
\item RQ3. What are the tactics to coordinate their acts?
\end{itemize}

\subsection{Expert insights}
We conducted two rounds of collaborative mapping with the SME group to identify the types of malicious activities and the classification mechanisms currently available. The purpose was to develop an understanding of how characteristics are attributed to facilitate the identification of malicious users, their behavior, and the tactics deployed in the orchestrated spread of disinformation.
In this study, we connected SME knowledge and academic literature to derive a taxonomy of malicious actors and their activities in OSNs. The SME activities provided insights into aspects of malicious actors' characteristics and how their behavior over time offers clues to their social influence and engagement strategies, reiterating previous profiling and detection studies \cite{zhou2020surveyfakenews}. Additionally, the activity enabled a shift in classification from issues of intentionality (e.g., willingness to engage in malicious online behavior). Therefore, our categorization focused on the entities' characteristics, behaviors, and tactics for broadcasting disinformation.

\subsubsection{Collaborative mapping}
\begin{figure}[h]
  \centering
  \includegraphics[width=\linewidth]{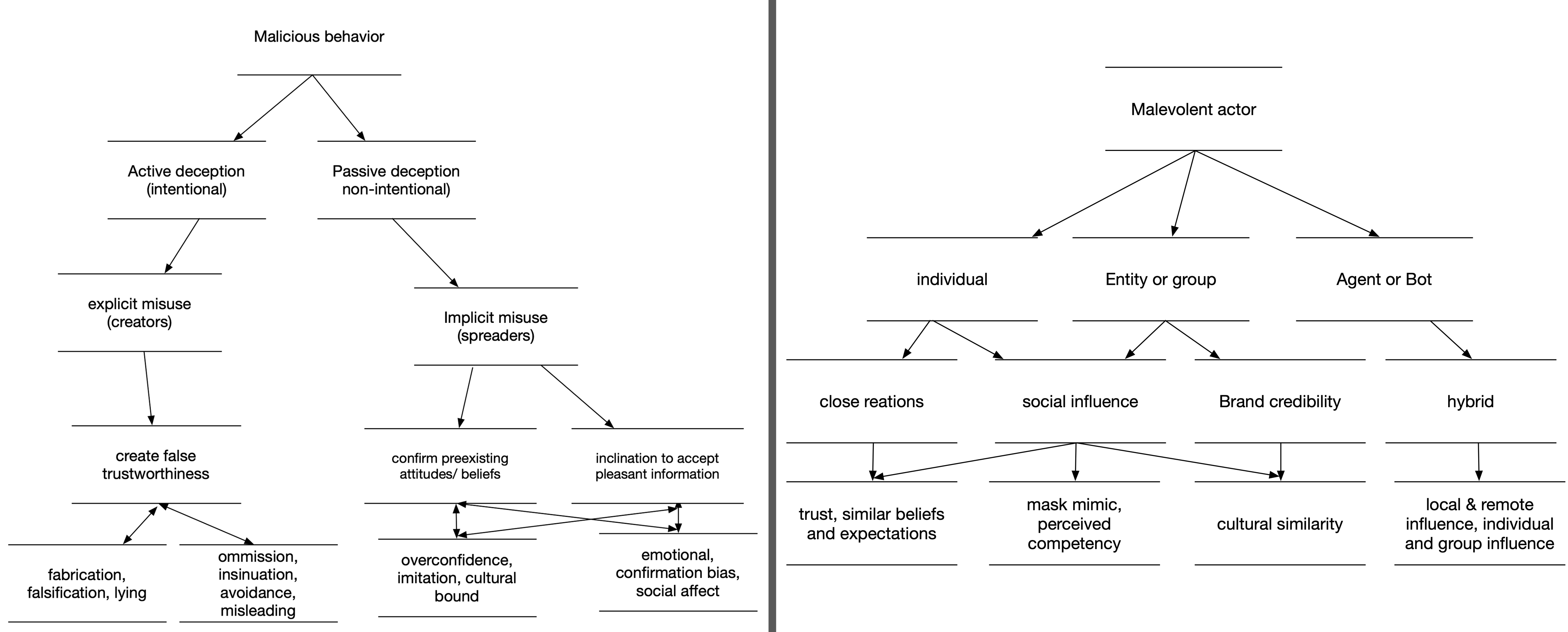}
  \caption{Participants mapped out the types of malicious actors and how they influence other members of OSN}
  \label{fig:colab}
  \Description{Groupings and classification of malicious actors focused on intentionality}
\end{figure}

The first round of collaborative activities with SMS enabled us to determine the angle our taxonomy should take on the investigated problem. Engaged in a joint activity, the SME group decided on the classification goals and agreed to reference the ending conditions based on the research questions proposed in our study. As shown in Figure~\ref{fig:colab}, the outcome of initial discussions was centered around behavior and the actors' intentionality. Participants discussed possible characteristics that could facilitate the densification of malicious users, their behavior, and the tactics used to orchestrate the spread of disinformation.

The participatory activity enabled a deeper understanding of how the behaviors and strategies of different types of malicious actors are incorporated into various disinformation campaigns. The discussion, which also unfolded around the "intention" factor in the spread of disinformation, allowed us to further refine the scope of the taxonomy, decentralizing it from issues of intentionality (e.g., willingness to engage in malicious online behavior).

Based on the results of the group activities, we classified malicious actors into two categories: creators, who create content and use specific mechanisms to foster credibility and engagement, and spreaders, who disseminate malicious content, whether intentionally or not. 
As noted in \cite{lesher2022disentangling}, these definitions also support a typology of engagement in disseminating false and misleading content, as well as how these entities forge trust and influence while interacting on social media platforms. Hence, our taxonomy focused on characteristics and patterns of behaviors that could suffice for the classification.

\subsection{Literature Review}
To complement the concepts gathered during the collaborative mapping, we conducted a review of the academic literature, focusing on peer-reviewed articles. The articles were searched via institutional databases using the Web of Science (N=81) and EBSCO (N=182) libraries, with constraints for publication year (i.e., 2013-2023), and English language. The search strings encircle three main dimensions of the problem our study meant to examine: Malicious Actors (i.e., "Malicious Actors" OR "Malicious Entities" OR "Malicious Users"); Online Social Networks ("Online Social Networks" OR "Social Media" OR "Social Media Platform" OR "Social Network Services"); and Disinformation ("Disinformation" OR "False Information" OR "Deceit Content"). 

After duplicates were removed (N=106), the remaining articles were assessed by abstract following the list of criteria for inclusion, framed around three factors: a) definitions, characteristics, and/or classifications of malicious actors; b) malicious actors' behaviors and actions; c) tactics and strategies for creating or spreading malicious content. A total of 39 articles were included in the full-text review, and data extraction was performed by four independent researchers who received a table listing the relevant information to be extracted from the articles. Following a standardized data extraction procedure, reviewers took notes on the following: author(s), reference, study's main point, references to machine learning models, nomenclature, and definitions of malicious actors, roles in disinformation spreading (creator, spreader, ambiguous), behaviors, actions, and strategies.

\subsubsection{Designing the classification}

We used a mix of tools (i.e., spreadsheets, physical and digital whiteboards) to analyze the dataset resulting from the juxtaposition of expert insights and academic literature. Following the principles of axial coding \cite{Vollstedt_Rezat_2019}, we used a dataset containing concepts and their definitions to examine the relationships between categories that emerged from the SME's discussion and academic concepts developed during the open coding process of the selected literature. Sequentially, we gathered all concepts to classify individuals or groups involved in the campaign, whether in planning, execution, or participation. 
Inspired by previous research that supports the use of characteristics of the malicious user's profiles for grouping the types of actors in a binary scheme, such as bots \textit{vs} real, creators \textit{vs} spreaders \cite{fire2014osnthreats}, we started by separating the malicious actors by their roles and features, following a strategy previously presented by \cite{patel2020human, lesher2022disentangling}. As mentioned by \cite{lesher2022disentangling}, these definitions also support a typology of engagement in disseminating false and misleading content and of how these entities build trust and influence while interacting on social media platforms. Also, we mapped the concepts retrieved from the literature, grouping the definitions according to the descriptions in the reviewed articles. Such a stage was oriented by the examination of the RQs. Lastly, while examining the manipulation tactics, we devise a set of categories to encapsulate the strategic plan of action of a disinformation campaign. Previous studies on media manipulation found common disinformation campaign tactics deployed across platforms \cite{Acker_Donovan_2019}.

\section{Results}

The approach taken in developing the taxonomy aims to address the issues of context-specific classifications previously identified in the literature. To do so, we developed three components to analyze the characteristics and activities of malicious actors, each addressing one of our research questions.

\subsection{Who are the malicious actors?}

\begin{figure}[h]
  \centering
  \includegraphics[width=\linewidth]{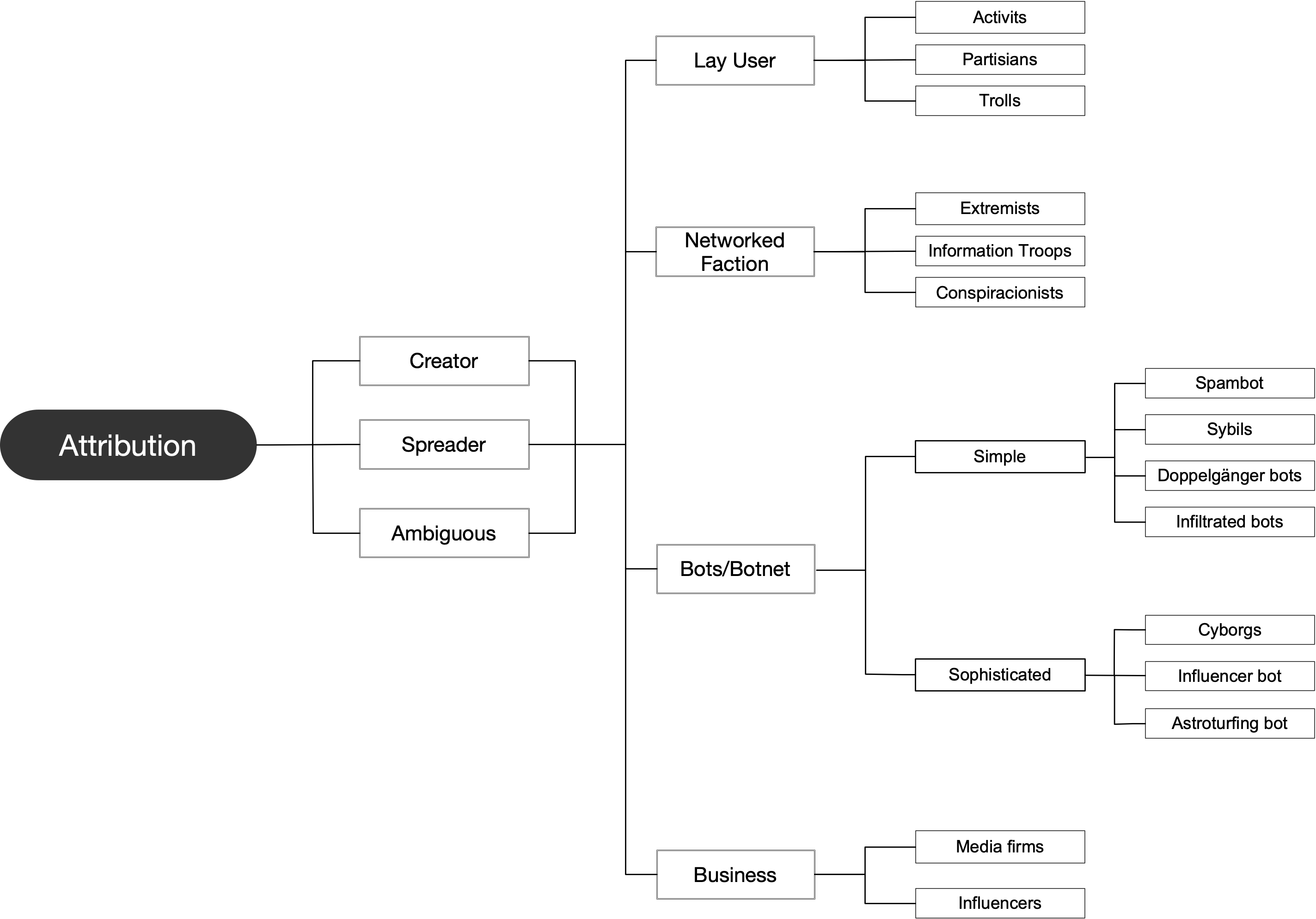}
  \caption{The general attribution provides a four-layer analysis for classification of malicious actors' roles and characteristics}
  \label{fig:attribution}
  \Description{The general attribution provides a four-layer analysis for classification of malicious actors' roles and characteristics}
\end{figure}

Regarding their nature, we classify malicious actors as a)creators, users who create content and use certain mechanisms to create credibility and engagement, b)spreaders, users who disseminate malicious content, with or without intention, and c)ambiguous, which refers to actors that function in dual roles by taking part in both the generation and dissemination of disinformation. Figure \ref{fig:attribution} presents the classification used in our taxonomy. 
On the second and third levels, the classification enables assessment of the actors' more detailed features. First, for analyzing the distinction between lay users and third-party customers, as recommended by \cite{Acker_Donovan_2019}. Later, ensuring that a categorical classification allows the selection of multiple low-level attributes provides additional dimensions for analysis.

\subsection{What do malicious actors do?}

\begin{figure}[h]
  \centering
  \includegraphics[width=12cm]{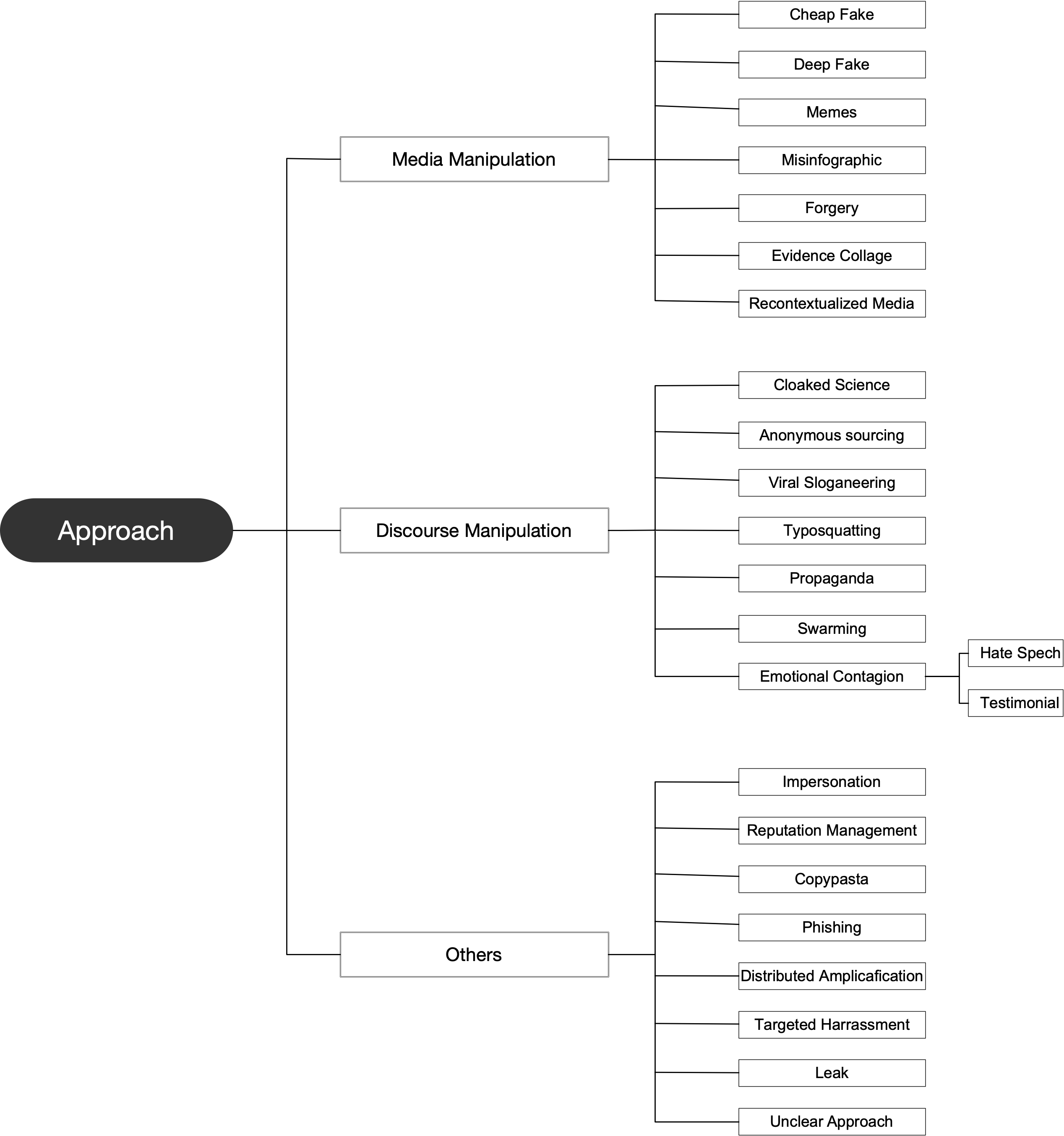}
  \caption{The approach classification allows differentiating the kind of manipulation done in the malicious content, encompassing the fusion of media manipulation tactics and computational tactics}
  \label{fig:approach}
  \Description{Approach}
\end{figure}

In the second dimension of our taxonomy, (B) Approach, we categorized a range of actions that malicious actors can utilize to promote a disinformation campaign. Figure \ref{fig:approach} presents the classification used in our taxonomy. This dimension encompasses three groups and 22 distinct categories, providing an option to define an unclear approach. The categorical classification allows multiple selection within a list of mechanisms employed to persuade, deceive, and create false credibility.

\subsection{What are the tactics to coordinate their acts?}

\begin{figure}[h]
  \centering
  \includegraphics[width=8cm]{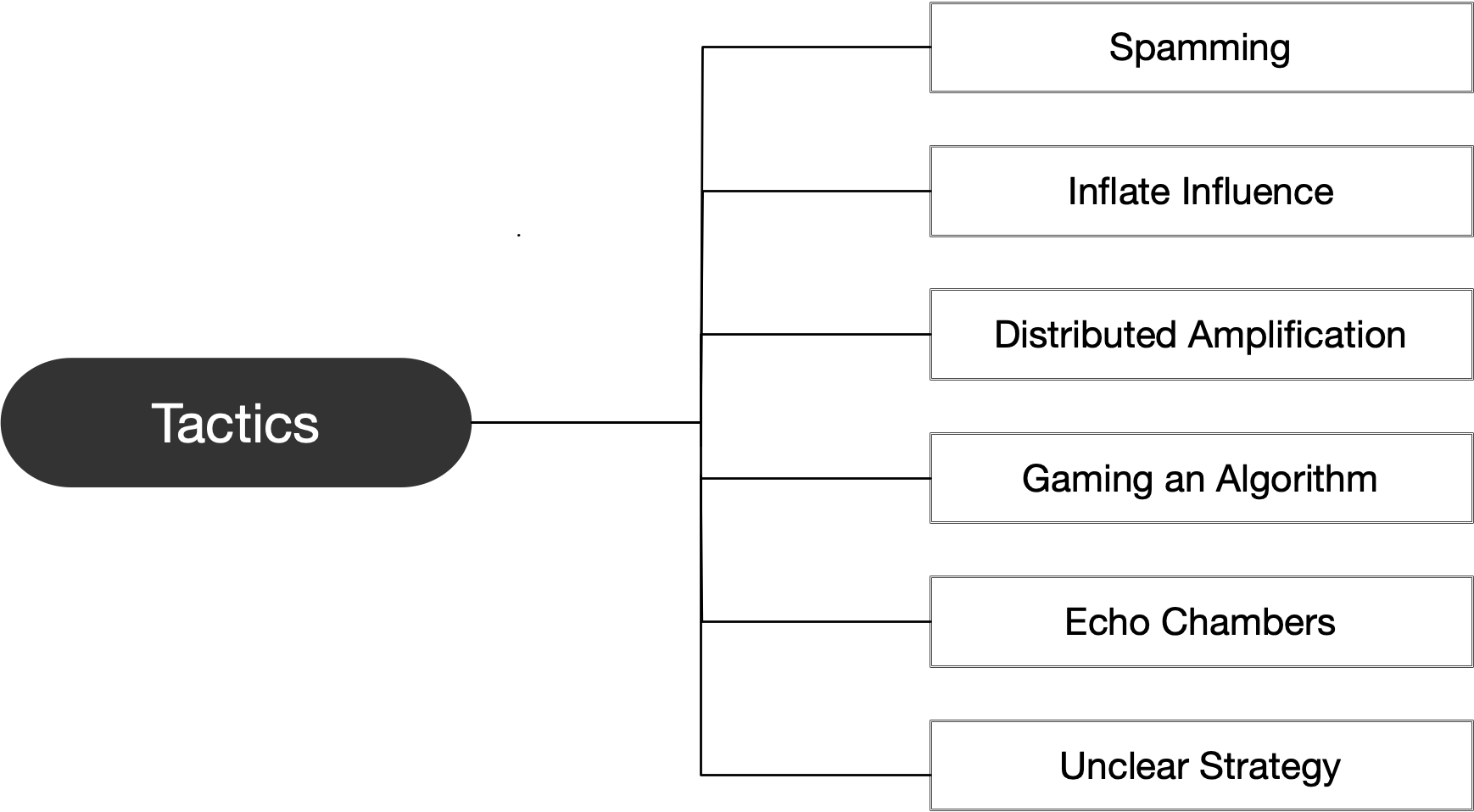}
  \caption{Six different strategies were identified as possible dissemination tactics}
  \label{fig:tactics}
  \Description{Six different strategies were identified as possible dissemination tactics}
\end{figure}

Finally, Tactics categorize the overarching plan of action used to accomplish the objectives of the disinformation campaign, as shown in Figure \ref{fig:tactics}. This dimension comprises six items that delineate strategies documented in the literature. Multiple selections are possible when assessing this dimension of malicious activities.

\section{Applying the Taxonomy}

Following the taxonomy development phase, a case study was conducted to assess the suitability of our taxonomy in a real-world context and to derive profiles from the categorized material. The annotation exercise was conducted to test the taxonomy as a tool for classifying malicious content retrieved from a targeted Telegram channel. A labeling software, designed by project partners, was used to facilitate the task and establish a standard across the research groups involved.  
The following sections present the process and discuss the opportunities and drawbacks we encountered. 

\subsection{Taxonomy refinement \& Label studio}

Once the taxonomy was defined, the challenge was to implement it in an integrated manner across the five countries and six institutions involved in the project, while accounting for both data privacy and data volume, as many messages contained images and videos in addition to text. The full version of the LabelStudio platform was selected—an open-source platform widely adopted for annotation tasks of this kind—and connected to the project’s proprietary database to prevent any data from being stored on the company’s servers. In this way, despite working with public and accessible data, no information was shared with the platform, thereby preserving data privacy.

\begin{figure}[h]
  \centering
  \includegraphics[width=\linewidth]{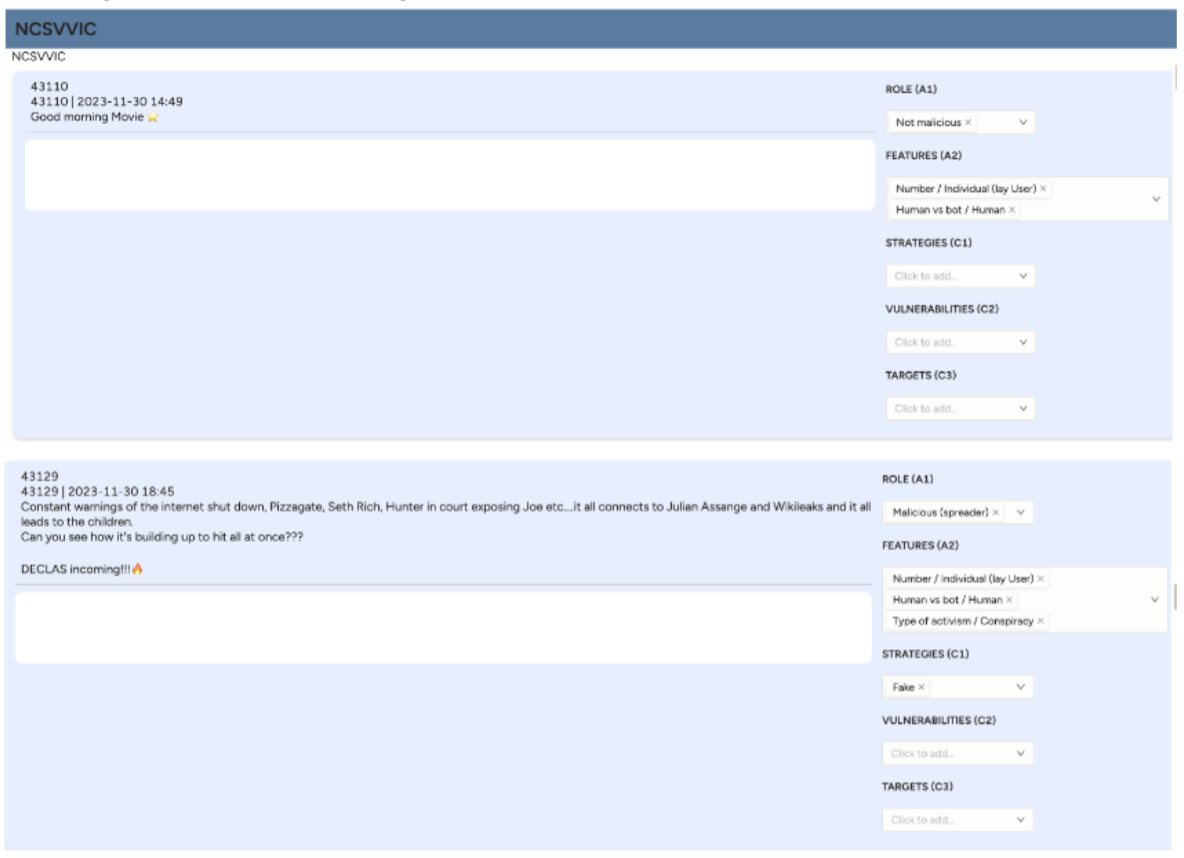}
  \caption{Example of two messages and the visualization of their annotations}
  \label{fig:label}
  \Description{Example of two messages and the visualization of their annotations}
\end{figure}

From a total of 6,805,626 messages extracted from 491 downloaded channels, an initial pilot study was conducted to evaluate the platform’s functionality and to identify limitations and potential improvements to the taxonomy. Subsequently, the last 100 messages from the selected channels were annotated due to platform issues and cost considerations, as detailed in the next section. To ensure adequate contextual understanding, annotation was carried out by teams from each participating country. Each message was independently annotated by three annotators in order to obtain a reliable measure of inter-annotator agreement (IAA) and to assess the overall reliability of the annotation process. Figure \ref{fig:label} presents two examples of the LabelStudio interface, showing the annotations assigned to each message on the image's right-hand side.

Although using a platform connected to the project’s databases introduced additional complexity in taxonomy implementation, connectivity, and storage, it also provided several important advantages. All teams were able to access shared information, and the consistency of annotations across different annotators could be systematically monitored. Furthermore, all data was centrally stored in a single database with automatic backups, thereby avoiding potential compatibility issues associated with local deployments of this tool or others. Finally, the use of an open platform and the storage of data in an open format (JSON) facilitated the straightforward export of results for subsequent analyses.

\subsection{\textbf{Case Study: Profiling Malicious Activities in Telegram Channels}}

To conclude our study, we evaluated the taxonomy's applicability to annotating malicious content collected from Telegram channels. The case study focused on malicious actors within the context of migration, aiming to apply a taxonomy on such actors. Channels that produced anti-migration discourse frequently exhibited profiles associated with politically extreme content. 

During the annotation process, annotators generally analyzed overall channel behavior to form an impression, as content-based labeling required this broader perspective. Addressing messages individually, without this contextual analysis, was often insufficient. The initial contribution of taxonomy to the content annotation process extended beyond the classification of ‘misinformation’ by enabling the dissection of content according to intent, target, strategy, and tactic. Reading through each message, the experts classified the type of actor, behavior, and tactics identified. 

The data were independently annotated, and the experts collected notes on the taxonomy's suitability for the classified data, its limitations, and possible improvements. This systematic classification facilitated the identification of sub-narratives, including emotional mobilization, manipulation, fear-mongering, and enemy construction.

This classification also enabled the identification of similar rhetorical patterns in channels featuring anti-migration discourse, enhanced the understanding of framing processes, particularly those based on fear, and revealed patterns related to supporting discourses. The taxonomy significantly facilitated subsequent analytical processes, including the coding of emotional manipulation, examination of hate speech narrative construction, and demonstration of the systematic generation of moral panic.

\textbf{Reflections on Applying the Taxonomy}

The developed taxonomy was theoretically comprehensive and included numerous useful subdivisions. However, certain features of the Telegram platform presented challenges during implementation. The primary issue involved ambiguous content. Some channels disseminated genuine news that did not contain misinformation, yet channels systematically producing and disseminating hate speech could frame such news within hateful narratives. As a result, while the content itself might be harmless, the framing process, selective sharing, and manipulative titling could create a harmful context from otherwise accurate news. This finding underscores that the definition of ‘malicious content’ cannot rely solely on the content itself and indicates that the taxonomy requires comprehensive warnings addressing this limitation.

Annotators encountered difficulties in some channels when applying the taxonomy, particularly with messages containing jargon specific to certain ideological groups. Such messages could represent parody, satire, or troll content, thereby raising concerns about contextual distortion.

Another platform-specific challenge arose from Telegram’s reliance on forwards and cross-channel circulation. The same visual or content could be recirculated across different channels with varying frames. This required careful attention to the distinctions between content production and circulation when applying the taxonomy.

\textbf{Reflection on implementation of the tagging}

The message annotation process spanned three weeks and, beyond its practical implementation, prompted a series of reflections on both the strengths and limitations of the proposed taxonomy. To this end, annotators were encouraged to record their doubts or observations directly on the platform via comments. These notes, along with insights from follow-up meetings, revealed several recurring challenges inherent in annotation strategies, which we present here to help refine similar efforts.

First, one of the key issues concerned the selection of messages for annotation. Given the variability in channel activity, ranging from a handful of messages to several thousand,  it was not feasible to annotate all the messages for all channels. Furthermore, technical tests showed that uploading the full content of these channels significantly hindered the platform's performance. Consequently, we opted to upload and annotate only the most recent 100 messages from each channel.

Second, this decision led some annotators to report difficulties in applying the taxonomy to the specific nature of content typically shared on Telegram. Messages were often highly fragmented or, conversely, unusually long and transmedia in nature, forming part of broader conversations that extended beyond the confines of a single channel. While the 100-message window was adequate in many instances, in other cases it proved insufficient for establishing a meaningful connection between individual messages and the taxonomy categories.

Third, and related to the previous points, there were instances in which annotators were aware that the general orientation of a channel was malicious — hence its inclusion in the study — yet, within the limited window of 100 messages, no content could be directly labeled as such. For example, the presence of legitimate news articles from reputable outlets, although malicious in a broader context, could not be classified as harmful in isolation.

Fourth, the lack of contextual information also stemmed from messages embedded within highly specialized or niche discourses, often unfamiliar to annotators. This posed a particular challenge, as distinguishing between satire, parody, jargon, and genuinely malicious content becomes difficult in the absence of sufficient contextual cues, a problem also noted in previous studies \cite{vasu2018fake, menczer2020information}. The involvement of multiple annotators (in this study, four researchers) facilitated the interpretation of contextual differences by establishing a minimum common ground.

Finally, another challenge of this approach was the frequent reuse and cross-posting of content across multiple channels. On Telegram, it is common for the same links or messages to be shared repeatedly or forwarded between different groups. This practice complicates annotation tasks and, more critically, risks introducing bias in model training if the broader context is not considered. The same message may or may not be malicious, depending on the discursive environment in which it appears.

\section{Discussion}

A new paradigm of interaction and communication is emerging with the pervasive presence of social media in our lives. Connected 24/7, individuals increasingly resort to OSNs to reach out to their communities, stay informed, and be heard. These online platforms provide space and infrastructure that enable individuals to share their thoughts, ideas, and creativity with large audiences, and even to form social communities around common interests. More than a network of users, OSN serves currently as a medium for communication, interaction, and collaboration. Nevertheless, the alarming and pervasive proliferation of deceptive content online raises concerns about how society receives and acts upon the online disinformation. 

Efforts to develop mechanisms to address malicious activity in OSNs draw on diverse approaches. However, the challenges in developing preemptive mechanisms to identify the activities of such individuals behind the dissemination of deceptive and fabricated content are manifold. The viral, widespread reach of disinformation campaigns on these platforms is a constant threat to institutions, extending beyond the constraints of a digital medium.  Recent advancements in large language models (LLMs) and deepfakes further complicate efforts to find long-term solutions to the issue. Moreover, online disinformation spreads entangle a complex network of users, behaviors, and dynamics in a fuzzy context. Moreover, the viral characteristics of deceptive content and amplification factors within online networks facilitate the spread of disinformation, creating several bottlenecks for the development of strategies to address such behavior across social media.

With that in mind, our study aimed to aid researchers and practitioners in developing generalizable, event-independent detection mechanisms to counter malicious activity in OSNs. As part of a broader project to develop machine-learning tools to detect and profile malicious actors, the authors examined mechanisms for categorizing them by their characteristics, activities, and coordination strategies. We adopted a collaborative and interdisciplinary approach, drawing on expertise in psychology, natural language processing, social network analysis, and human–computer interaction.

To answer RQ1, we proposed the general attribution classification presented in Figure 1. In doing so, we categorize malicious actors into Creators and Spreaders. This category is mutually exclusive (i.e., no actor could have two characteristics). Nonetheless, the annotators' ambivalence about the actors' roles led us to include a third item, Ambiguous, because it is often impossible to trace the origin of content on OSNs. Therefore, introducing a feature-level classification enabled us to describe various types of malicious actors.
Moreover, the sequential layers of the Attribution classification allow for deeper reflection on the actor's nature before proceeding to analyze their behavior. To demonstrate this, although the example does not refer to a malevolent OSN agent, we can consider the artificial-intelligence influencer Lil'Miquela, introduced in 2016 by a Los Angeles-based company. Assessing her attribution in our taxonomy will show that Miquela is a \textit{Creator>>Bots/Botnet>>Sophisticated>>Influencer bot}, but at the same time, was shown by \cite{roy2025_ai_influencers} analysis of the AI influencers and media firms, as Miquela is also the representant of a successful business, Miquela shouwld be classified as \textit{Creator>>Business>>Influencers}.

The Approach category describes the activities and the techniques used to deceive OSN users and spread disinformation, as presented in Figure 2. This dimension of the taxonomy provides three higher categories to differentiate the types of manipulation observed in malicious content, encompassing the fusion of media and computational tactics. Inside \textit{Others}, we classify extra approach mechanisms which include targeted harassment, reputation management, and \textit{copypasta} - an emerging category of disinformation spreading identified by \cite{Acker_Donovan_2019} in which blocks of text or screenshots are repeatedly (re)posted on social media, online forums, and comment sections. More, observing the Approach used by the malicious actors aims to account for the re-contextualization often seen in which content is neither deceptive nor malicious but becomes harmful within a specific channel or context.

Lastly, to address RQ3, we introduced a third dimension to our taxonomy, as shown in Figure 5. Our goal was to complement the identification of malicious activities by assessing possible tactics, such as comprehensive strategies deployed in disinformation campaigns. As in previous research \cite{alam2021survey-b92}, which advocates combining multiple information modalities, we highlight combining multiple dimensions of the taxonomy as an advantage for improving the classification of malicious content.

In this context, the taxonomy provides a robust foundation for content analysis. The attribution of different roles and manipulation mechanisms presented in this study also helps to clarify how malicious actors operationalize authentic and inauthentic behavior on OSN platforms. Despite these limitations, the taxonomy proved effective in training AI models to identify malicious actors in the context of the Russia–Ukraine conflict. It was also successfully applied—albeit to a lesser extent—to ongoing research on climate change misinformation on TikTok and antivaccination communities.

\subsection{Limitations}
Identifying and classifying malicious actors involves interpretive judgments that can be influenced by contextual biases or the limitations of available data. Additionally, broader validation across multiple scenarios, platforms, and geographic regions is necessary to confirm the robustness and utility of this malicious agents classification.

As online social OSNs evolve rapidly, the taxonomy may need continuous updates to stay relevant. Finally, translating these insights into practical design interventions involves additional challenges outside this study's scope. These challenges include platform-specific constraints and ethical trade-offs. Future research should address these limitations through longitudinal studies, particularly those examining cross-platform analyses. However, the classification process exhibited certain limitations. Some messages, although not a direct indication for measuring contextual harm, were not entirely sufficient. Another challenge involved determining intent; distinguishing between irony, mockery, trolling, and deliberate manipulation often required interpretation to ascertain the actor's true intent. Some messages, although not directly containing misinformation, could reinforce harmful narratives within the context, such as criminalization, which generalizes a single crime to an entire group.

Additionally, when Telegram messages were multi-layered and included visuals or videos alongside text, one-dimensional labeling proved inadequate. These factors increased the complexity and responsibility of the human annotator, making the process reliant on human interpretation. Although this reliance can be viewed as a constructive limitation in the interpretation of social events, it underscores the necessity of addressing these issues for future users. It's noteworthy that the research activities took place in 2023, before Telegram's policy changes \cite{jamali2024telegram} aimed at reducing illicit activities on the platform.

\section{Conclusion}

In this study, we designed and demonstrated the application of a taxonomy of malicious actors, their roles, and tactics for broadcasting disinformation on OSNanan. A multidisciplinary team joined efforts to provide a event-independent classification that focuses on the nature of the agents and their behavior. Online disinformation spread is a complex and imprecise issue that requires mechanisms developed through interdisciplinary approaches.

Disinformation spreading has been linked to several factors related to the OSN algorithm infrastructure \cite{Dourado2023}, user emotions \cite{Rastogi2022}, social and political polarization \cite{PastorGalindo2020}, network effects (influential spreaders) \cite{conti2017s}, lack of verification mechanisms \cite{Shao2018}, as well as user's perception of their own anonymity \cite{dupuis2020information}. Moreover, the amplification of disinformation is closely related to digital and information literacy and to the judgment that lay users make about what is shared and disseminated \cite{karlova2013social}. In addition to the impact of orchestrated disinformation campaigns, users are subjected to echo chambers \cite{conti2017s}. Also, to exert control and influence, malicious actors use information collected from users' past preferences to deliver targeted content recommendations and personalized search engine results (menczer2020information).

As our research demonstrated, the challenge is intensified by the improvement of malicious strategies deployed to evade detection mechanisms within social network communities and to avoid bans and other punishments described in social networks' guidelines. Even so, we believe the contributions of this study can aid researchers and practitioners in developing generalizable, event-independent detection mechanisms to counter malicious activity in OSNs.

The research and categorization employed in our taxonomy focused on the actors and their activities. While performing annotation and analysis at the message level across different channels' content, we observed that the taxonomy provides a robust foundation for content analysis. However, its application across different platforms underscores the need to consider each platform’s structural differences and requires annotators to exercise informed judgment. The involvement of multiple annotators (in this study, four researchers) facilitated the interpretation of contextual differences by establishing a minimum common ground.

Telegram is a messaging application that was selected as our case study as it was attractive to malicious actors for its lack of strict content moderation and channel features that allow users to broadcast messages to large, private groups, making it difficult for law enforcement and security researchers to monitor malicious activities \cite{Rogers2020}.
However, as demonstrated in our case study, the content posted in the flagged channels often employed approaches such as discourse or media manipulation, though stemming from different tactics. This way, as previous studies had demonstrated, malicious content is disseminated in an interconnected manner, in which repeated content is shared by actors and, even leading users to external platforms and websites spamming identical content \cite{Alom2018}, or tricking users by engagement-based ranking algorithms which create the appearance that some person or opinion is popular \cite{Yang2019}. 

Our taxonomy, nevertheless, provides an opportunity for multi-layered, multi-actor analysis, enabling the classification of content on Telegram channels according to dimensions such as emotional manipulation, disinformation, divisive rhetoric, and conspiracy narratives. Furthermore, the taxonomy provided substantial analytical evidence for specific patterns in anti-migration discourse. It facilitated the identification and demonstration of how emotional mobilization and anger production were instrumentalized, particularly when migrants were framed using arguments related to security, threat, economic burden, and cultural decline. This pattern contributed to a deeper understanding of the platform's alternative information ecosystem and supported the development of a holistic theoretical framework for the research. 

\begin{acks}
This work has been supported by the project ``Malicious Actors Profiling and Detection in Online Social Networks through Artificial Intelligence'' (MARTINI), grant CHIST-ERA-21-OSNEM-004, and the Estonian Research Council (TKA22209).
The author, Berta Chulvi, carried out this work as part of the research team at the Universitat Polit\`ecnica de Val\`encia.
\end{acks}

\bibliographystyle{ACM-Reference-Format}
\bibliography{references}

\end{document}